\newcommand{\eq}[1]{(\ref{#1})}
\newcommand{\kb}{k_{\rm B}}
\newcommand{\tw}{t_{\rm w}}
\newcommand{\Lo}{L_{0}}

\newcommand \be {\begin{equation}}
\newcommand \ee {\end{equation}}
\newcommand \beq {\begin{equation}}
\newcommand \eeq {\end{equation}}
\newcommand \bmat {\begin{displaymath}}
\newcommand \emat {\end{displaymath}}

\newcommand \qea {q_{\rm EA}}
\newcommand \qd {q_{\rm D}}
\newcommand \chiea {\chi_{\rm EA}}
\newcommand \chid {\chi_{\rm D}}

\newcommand{\figwidth}{3.375in}

\documentclass[aps,prl,twocolumn,superscriptaddress]{revtex4}

\usepackage{graphicx}% Include figure files

\begin{document}

\title[Short Title]{Anomalously Soft Droplets and Aging in Short-ranged Spin Glasses}

\author{Hajime Yoshino}
\email{yoshino@ess.sci.osaka-u.ac.jp}
\affiliation{
Department of Earth and Space Science, Faculty of Science, 
Osaka University, Toyonaka, Osaka 560-0043, Japan \\
}
\author{Koji Hukushima}
\email{hukusima@issp.u-tokyo.ac.jp}
\affiliation{
Institute for Solid State Physics, University~of~Tokyo, 5-1-5 Kashiwa-no-ha,
Kashiwa, Chiba 277-8581,\\
 Japan}
%\email{hukusima@issp.u-tokyo.ac.jp}
\author{Hajime Takayama}
\email{takayama@issp.u-tokyo.ac.jp}
\affiliation{
Institute for Solid State Physics, University~of~Tokyo, 5-1-5 Kashiwa-no-ha,
Kashiwa, Chiba 277-8581,\\
 Japan} 

\date{\today}

\begin{abstract}
We extend the standard droplet scaling theory for isothermal aging 
in spin glasses assuming that the effective stiffness constant of droplets 
as large as extended defects is vanishingly small.
A novel {\it dynamical} order parameter $\qd$  and
the associated dynamical susceptibility $\chid$ emerge.
Breaking of the fluctuation dissipation theorem takes place at $\qd$
well below the {\it equilibrium} Edwards-Anderson order parameter  
$\qea (> \qd)$ at which the time translational invariance
is strongly broken. The scenario is examined numerically in a 4 dimensional 
EA Ising spin glass model.
\end{abstract}

\maketitle

Recently aging in spin glasses 
\cite{APY-review,Saclay,Uppsala,Review-Rieger,BCKM} 
has attracted renewed interest both in experimental, numerical 
and theoretical studies.
The phenomena provide very interesting and almost the only realistic way
to explore large scale, non-trivial properties of glassy phases.
A major theoretical progress was the development 
of dynamical mean-field
theories (MFT) for spin glasses and related systems \cite{MFT} which lead
to findings of some novel view points in glassy dynamics such as
the concept of effective temperature \cite{CKP97}. 
However, the MFT themselves
do not provide insights into what will become important in 
realistic finite dimensional systems, such as nucleation processes.

The droplet theory \cite{BM,FH1,FH2} developed
a scaling theory assuming droplet 
excitations as basic nucleation processes.
It provides a perspective for equilibrium and
dynamical properties of finite dimensional spin glasses. 
Though it was developed more than a decade 
ago, many of its predictions have remained to be clarified. 

Recently, anomalous low energy and large scale excitations 
were found \cite{LEE} in active studies of spin-glass models at $T=0$.
In the present letter, we develop a refined
scenario for isothermal aging in spin glasses where such
anomalous low energy excitations play a significant role. 
Most importantly, it explains
the fundamental experimental observation that
the field cooled (FC) susceptibility is larger than 
the zero field cooled (ZFC) susceptibility \cite{Uppsala,Saclay}.
Such an anomaly was reproduced successfully in the dynamical MFT
but not in the conventional droplet theory.
To simplify notations, we discuss systems 
of $N$ Ising spins $S_{i}=\pm 1$ ($i=1,\ldots,N$) 
in a $d$-dimensional space
coupled by short-ranged interactions
of energy scale $J$ with random signs with no ferromagnetic or 
anti-ferromagnetic bias.
Extension to other types of spin glasses can be done straightforwardly.

%\section{Softening of Droplets}

\newcommand{\Lmax}{L_{\rm m}}
\renewcommand{\Lo}{L_{0}}
\newcommand{\rhoo}{\tilde{\rho}(0)}
\newcommand \FtypLR {F_{L,R}^{\rm typ}}
\newcommand \Ueff {\Upsilon_{\rm eff}}

Fisher and Huse \cite{FH2} noticed that droplet excitations
can be {\it softened} in the presence of a {\it frozen-in} defect 
or domain wall compared with ideal equilibrium with no
extended defects. This is because droplets which touch 
the defects can reduce the excitation gap compared with those 
in equilibrium. 
Let us consider system with a frozen-in defect of size $R$:  
a large  droplet of size $R$ is flipped with respect to $\Gamma$,
which is a ground state of an infinite system, and then it is frozen. 
The typical free-energy gap $\FtypLR$ of a smaller 
droplet of size $L$ in the interior of the frozen-in 
defect is expected to scale as
\be
\FtypLR = \Upsilon_{\rm eff}[L/R](L/L_0)^\theta   \qquad L < R
\label{eqn:effU-general}
\ee
with $\Lo$ being a certain microscopic length scale and 
$\Ueff[L/R]$ being an {\it effective stiffness} 
which is only a function of the ratio $y=L/R$.
We assume as usual that the probability distribution of the free-energy gap
$F_{L,R}$ follows the scaling form
$\rho(F_{L,R})dF_{L,R} = 
\tilde{\rho}(F_{L,R}/\FtypLR) dF_{L,R}/\FtypLR$ with non-vanishing amplitude
at the origin $\rhoo > 0$ which allows marginal droplets.

For $y \ll 1$, $\Ueff[y]$ will decrease with $y$ as~\cite{FH2}
\begin{equation}
\Upsilon_{\rm eff}[y]/ \Upsilon=1-c_{v} y^{d-\theta} \qquad 
\mbox{for} \qquad   y \ll 1.
\label{eqn:effU-small}
\end{equation}
Here $\Upsilon$ is the original stiffness constant $\Upsilon=\Ueff(0)$. 
A basic conjecture, on which our new scenario based, is that 
 at the other limit $y \sim 1$ the effective stiffness vanishes as,
\begin{equation}
\Upsilon_{\rm eff}[y]/\Upsilon \sim (1-y)^{\alpha}
\qquad  y \sim 1
\label{eqn:effU-large}
\end{equation}
with  $ 0 < \alpha < 1$ being an unknown exponent, and that the lower
bound for $\FtypLR$ should be of order $J$, say $ F_{0}$. 

In spin glasses of finite sizes $R$, it is likely that 
the existence of boundaries will intrinsically induce certain defects
as compared with infinite systems \cite{NS,M99,KYT}. 
Then droplet excitations as large as the system size itself $L \sim R$
may be anomalously soft as conjectured above.
Such an anomaly is indeed found in recent studies \cite{LEE}.
It explains the apparently non-trivial overlap distribution
function $P(q)$ found in
numerous numerical studies of finite size systems \cite{low-temp-MC}. 
Although there new exponent $\theta'=0$ was conjectured \cite{LEE}, 
we consider it better to attribute this to the zero stiffness constant 
$\Ueff[1]=0$ as in (\ref{eqn:effU-large}). The stiffness exponent
$\theta > 0$, on the other hand, is associated with a defect in $\Gamma$
as we adopt in (\ref{eqn:effU-general}).

To explain the consequence of our conjecture above introduced, 
let us consider thermal fluctuations and magnetic linear responses of
droplet excitations of size $L$ within the frozen-in defect at
scale $R$. Each droplet excitation 
will induce a random change of the magnetization of order
$M_{L} \sim m\sqrt{(L/\Lo)^{d}}$ where
$m$ is the average magnetic moment within a volume of $\Lo$. 
The thermal fluctuation can be measured by an
order parameter $q=N_{d}^{-1}\sum_{i}\langle S_{i}\rangle^{2}$ where the sum
runs over sites in the interior of the frozen-in defect which
contains  $N_{d} \propto (R/L_{0})^{d}$ spins. 
The magnetic linear response by weak external magnetic field $h$ 
is measured by a linear susceptibility 
$\chi=N_{d}^{-1}\sum_{i}\langle S_{i}\rangle_h/h$ where $\sum_{i}\langle
S_{i}\rangle_h$ is the induced magnetization by the field.
In the absence of  any droplet excitations,  $q=1$ holds due to the
normalization of spins. The reduction
from $1$ due to a droplet excitation at scale $L$ is of order
$M^{2}_{L} \rhoo(\kb T/\FtypLR)$ with $\kb$ being the Boltzmann constant.
Correspondingly, the induced magnetization of droplets by $h$ at scale
$L$ is of order $M_{L} \rhoo(hM_{L}/\FtypLR)$.

Let us construct a toy droplet model 
\cite{FH1} defined on logarithmically separated shells of length scales 
$L/\Lo=a^{k} < R/\Lo$ with $a>1$ and $0 \leq k \leq n_{2} \leq n_{1}$ 
where $R/\Lo=a^{n_1}$ and $\Lmax/\Lo=a^{n_2}$.
For each shell an {\it optimal } droplet is assigned whose 
free-energy gap is minimized within the shell. 
Then $\FtypLR$ will be of order $F_{0}$ at the shell $k=n_{1}$
so that we have
$\FtypLR=\FtypLR(1-\delta_{k,n_1})+F_{0}\delta_{k,n_1}$. 
For simplicity, droplets at
different scales are assumed to be independent from each other.
Replacing the sum  $\sum_{k=0}^{n_2}$ by an integral 
$\int_{\Lo}^{\Lmax}dL/L$ we obtain, 
\begin{eqnarray}
&&  1-q (\Lmax,R)  =  \kb T\chi(\Lmax,R) 
   =    \rhoo m^{2} \kb T \times \nonumber \\
&&  \int_{L_{0}}^{\Lmax}   \frac{d L}{L} \left[  
  \frac{1-\Delta_{a} (\ln (L/R)) }{\FtypLR}
 +
\frac{\Delta_{a} (\ln (L/R))}{F_{0}}
\right]    \hspace*{.5cm}
\label{eq-q-chi-lr}.
\end{eqnarray}
where $\Delta_{a}(z)$ is a {\it pseudo} $\delta$-function of 
width $\ln a$ \cite{note}.
Note that the fluctuation dissipation theorem (FDT) is satisfied.

It is useful to consider assymptotic behaviour at large sizes 
$R/\Lo \gg 1$ with the ratio 
$x=\Lmax/R$ being fixed. We obtain for $0 < x \leq 1$,
\be 
 q (xR,R) = \qea 
+ \frac{\rhoo m^{2}\kb T}{\Upsilon(R/\Lo)^{\theta}} A(x)
-\frac{\rhoo m^{2}\kb T}{F_{0}}\Theta_{a} (\ln x)
\label{eq-qxRR}
\ee
with $\Theta_{a}(z)$ being a  {\it pseudo} step-function of width $\ln a$
 \cite{note} and $
A(x)=
\int_{x}^{\infty}dy y^{-1-\theta} 
-\int_{0}^{x}dyy^{-1-{\theta}} 
(\Upsilon/\Upsilon_{\rm eff}[y](1-\Delta_{a} (\ln y))-1).
$
Note that the second integral  converges 
because of \eq{eqn:effU-small} and 
the inequality $\theta < (d-1)/2$ \cite{FH1}.
Here $q_{\rm EA}$ is the usual Edwards-Anderson (EA) order parameter
evaluated in the absence of any extended defects as
\be
q_{\rm EA} \equiv\lim_{\Lmax \to \infty}\lim_{R \to \infty}q(\Lmax,R)
=1-c \rhoo m^{2}\frac{\kb T}{\Upsilon}
\ee 
with  $c=\int_{1}^{\infty}dyy^{-1-\theta}$. The order of the limits is
crucial. 
Thus as far as $0 < x < 1$, the order parameter converges to
the {\it equilibrium} EA order parameter $\qea$ in the large size limit.
The associated {\it equilibrium}  susceptibility $\chiea$ is
defined as $\kb T\chiea \equiv 1-\qea$. 

In the intriguing case $x \sim 1$, $A(x)$ will remain finite
as far as $0 < \alpha < 1$. At $x \sim 1$ 
the last term of  \eq{eq-qxRR} contributes and we obtain,
\begin{eqnarray}
q(R,R) = \qd+  \rhoo m^{2}\frac{\kb T}{\Upsilon(R/\Lo)^{\theta}} A(1),
\label{eq-q-cross}
\end{eqnarray}
where we have defined the {\it dynamical order parameter} 
\be
q_{D}\equiv\lim_{R \to \infty}q(R,R) =\qea - 
\rhoo m^{2}\frac{\kb T}{F_{0}}.
\ee
This is one of the main results of the present work. 
Naturally, we can define 
the associated dynamical linear susceptibility $\chid$ as
$\kb T\chi \equiv 1-q_{D}$. As we discuss below 
$\qd$ and $\chid$ play significantly important role
in the dynamical observables of aging.

%\section{Scaling of Two-Time Quantities}

\begin{figure}[h]
\resizebox{\figwidth}{5cm}{\includegraphics{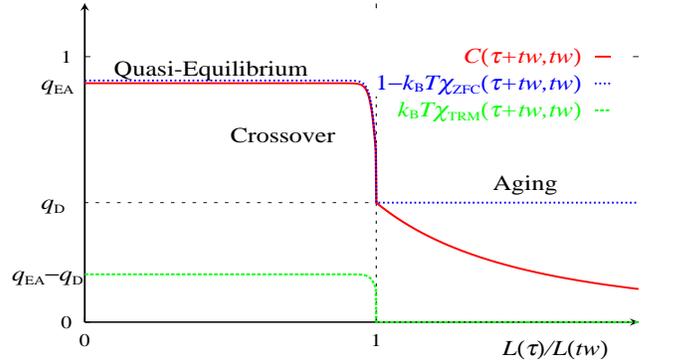}}
 \caption{Different asymptotic regimes of the two-time quantities.
Here the asymptotic limit $L(\tw) \to \infty$ 
is considered with the ratio $x=L(\tau)/L(\tw)$ being 
fixed. }
\label{fig:c-conjecture}
\end{figure}

We are ready to consider a simple isothermal aging protocol: 
the temperature is quenched \cite{cooling rate}
at time $t=0$ to a temperature $T$ below
its transition temperature $T_{c}$ from above $T_{c}$ and 
the relaxational dynamics i.e., aging is monitored.
The dynamical magnetic susceptibility
is generally defined as $\chi(t,\tw)=N^{-1}\langle  M(t) \rangle/h$
where $ \langle  M(t) \rangle $ is the total magnetization measured
at time $t$ and field $h$. 
In the thermoremanent 
magnetization (TRM) measurement the field is switched on 
for a waiting time $\tw$ and then cut off: $h(t)=h\theta(\tw-t)$.
On the other hand, in the ZFC magnetization measurement, 
the field is switched on after the waiting time: $h(t)=h\theta(t-\tw)$.
We denote the susceptibilities of these protocols as 
$\chi_{\rm TRM}(t,\tw)$ and $\chi_{\rm ZFC}(t,\tw)$.
Another important quantity is the magnetic autocorrelation function,
$C(t,\tw)=N^{-1}\langle M(t)M(\tw)\rangle=N^{-1} \sum_{i}\langle
S_i(t)S_i(\tw)\rangle$. The last equation follows
in the absence of ferromagnetic or anti-ferromagnetic bias.

Within the droplet picture, aging proceeds by coarsening
of domain walls \cite{FH2} as usual phase ordering 
processes \cite{B94}. Following \cite{FH2}, the whole 
process may be divided into {\it epoch}s such that the typical size of the
domain is $\Lo,a\Lo,a^{2}\Lo,\ldots, a^{n}\Lo,\ldots$.
At each epoch, droplets of various sizes up to that of the domain 
can be thermally
activated or polarized by the magnetic field. Importantly the
domain wall serves as the frozen-in extended defect for the 
droplets and reduces their stiffness constant.

In the droplet picture \cite{FH2} all dynamical processes, including both 
the domain growth and droplet excitations in the interior of 
the domains, are thermally activated processes which yield
a logarithmic growth law of the time dependent length scale $L(t)$.
%At large enough 
%time scales, the typical size of a nucleation which takes place 
%at time  $t$ is $L(t) \sim \Lo((\kb T/\Delta(T))\ln(t/\tau_0(T)))^{1/\psi}$ 
%with $\tau_0(T)$ being a  unit time scale for the 
%activated processes, $\psi$ the exponent of free-energy barrier, 
%and $\Delta(T)$ the energy scale of free-energy
%barrier. 
In practice, empirical power laws with 
temperature dependent exponent work as well\cite{KYT,Uppsala-AC}.
However, the logarithmic growth law is supported by a recent 
experiment \cite{Uppsala-AC} where the effects of 
critical fluctuations are considered (see also\cite{HYT,Saclay-AC,YHT,BB}).
In the following, we focus on scaling properties of the two 
time quantities as functions of the dynamical length scale
rather than on the growth law itself.

As far as $L(\tau)\leq L(\tw)$, 
the autocorrelation function and the ZFC  susceptibility 
are related to the generalized order parameter 
and linear susceptibility defined in
\eq{eq-q-chi-lr} as $C(\tau+\tw,\tw)=q(L(\tau),L(\tw))$ and 
$\kb T \chi_{\rm ZFC}(\tau+\tw,\tw)=1-C(\tau+\tw,\tw)$ 
satisfying the FDT.
Especially, the equilibrium correlation and response are 
obtained by taking the limit $L(\tw) \to \infty$ with fixed $L(\tau)$.
In this special limit one finds \cite{FH2},
$
C_{\rm eq}(\tau) 
=q_{\rm EA} +  c \rhoo m^{2}(\kb T/\Upsilon)(L(\tau)/L_{0})^{-\theta}
$
and
\be
\chi_{\rm eq}(\tau) \equiv \lim_{\tw \to \infty} \chi_{\rm ZFC}(\tau+\tw,\tw)
=\chi_{\rm EA} -  c \frac{\rhoo m^{2}}{\Upsilon(L(\tau)/L_{0})^{\theta}}.
\label{eq-chi-eq}
\ee

In the quasi-equilibrium regime $L(\tau) < L(\tw)$ waiting 
time dependence or violation of TTI (Time translational invariance)
is present in a {\it weak} manner as correction terms
to the ideal equilibrium behavior \eq{eq-chi-eq}.
The scaling form of the correction terms can be
found using \eq{eqn:effU-small} which is relevant for relaxation 
of AC susceptibilities \cite{FH2,Uppsala-AC}.
Actually it allows one to set up numerical extrapolations
to obtain the ideal equilibrium behavior \cite{KYT,HYT,YHT}.
In the limit $L(\tw) \to \infty$ with 
$x=L(\tau)/L(\tw)<1$ being flexed, we find
$C \to q_{\rm EA}$ and $\chi_{\rm ZFC} \to \chiea$ (see 
Fig. \ref{fig:c-conjecture}).
Much {\it stronger} violation of TTI shows up below $\qea$.

In the crossover regime $L(\tau) \sim L(\tw)$, 
the anomalously soft droplets as large 
as the size of the domain $L(\tw)$ are also needed to be taken into account.
From \eq{eq-q-cross} we immediately find,
$
 C(\tau+\tw,\tw)|_{L(\tau) \sim L(\tw)}
 \sim  q_{\rm D} + A(1) \rhoo m^{2}(\kb T/\Upsilon) (L(\tw)/\Lo)^{-\theta} 
$
where $\qd$ is the novel dynamical order parameter we introduced above. 
Thus as a function of $x$, there should be
a vertical drop from $\qea$ to $\qd$ at 
$x \sim 1$ in the asymptotic limit $L(\tw) \to \infty$. 
Correspondingly, the ZFC susceptibility
should jump up vertically from $\chiea$ to $\chid$ at $x \sim 1$
in the same asymptotic limit (see Fig. \ref{fig:c-conjecture}).
Such an abruptness will be absent as functions of $\tau/\tw$.
Indeed it is well known in  experiments \cite{Uppsala}
that the so called relaxation rate 
$S(t)=d \chi_{\rm ZFC}(\tau+\tw,\tw)/d \log (\tau/\tau_{0})$ 
has a pronounced peak at around $\tau \sim \tw$.

In the aging regime $L(\tau) \sim L(t) > L(\tw)$ we expect,
\be
C(t,\tw) \sim  q_{\rm D}\tilde{C}\left( \frac{L(t)}{L(\tw)} \right). 
\label{eq-scale-c-aging}
\ee
Here, following Fisher and Huse \cite{FH2}, we assume
the scaling function $\tilde{C}(x)$ is related to a probability 
$P_{\rm s}(L(t)/L(\tw))$
that a given spin belongs to the same domain 
at the two different {\it epochs} 
characterized by $L(t)$ and $L(\tw)$.
It satisfies $\tilde{C}(1)=1$ and 
$\tilde{C}(x)\sim x^{-\lambda}$ at $x\gg 1$
with $\lambda$ being in the range $d/2 < \lambda < d$.
It should be emphasized that, 
in contrast to usual domain growth process \cite{B94},
we put the dynamical order parameter $\qd$ 
rather than $\qea$ as the amplitude of $C(t,\tw)$ since we 
obtained $C(t,\tw)|_{x\sim 1} \sim \qd$ above.

The ZFC  susceptibility in the aging regime becomes,
\begin{eqnarray}
\chi_{\rm ZFC}(t,\tw) 
&& \sim  \chi(L(\tau),L(t)) \nonumber \\
&& +c'' \int_{L(\tw)}^{L(t)}\frac{dL}{L}
 \frac{\rhoo m^{2}}{\Upsilon(L/L_{0})^{\theta}}
\tilde{C}\left(\frac{L(t)}{L}\right) 
\label{eq-chizfc-aging}
\end{eqnarray}
where $c''$ is a numerical constant.
The first term is due to the response of 
the droplets equilibrated within the temporal domain of size $L(t)$
(see \eq{eq-q-chi-lr}).
The second term represents contributions from 
the magnetizations (per spin) of order 
$\rhoo h m^{2}/\Upsilon(L/\Lo)^{\theta}$ 
due to droplets as large as $L$ which are induced by the field in 
the {\it past  epochs}, say around $t$ with $L(t)=L$, and are 
depolarized by further domain growth. We also note that this term 
but with the integration from $\Lo$ to $L(\tw)$ yields the 
TRM susceptibility written as~\cite{FH2} 
$
 \chi_{\rm TRM}(t=\tau+\tw,\tw) 
 \sim  c_{\rm nst} \rhoo m^{2}/\Upsilon(L(\tw)/L_{0})^{-\theta}
(L(t)/L(\tw))^{-\lambda}
%\label{eq-chitrm-aging}
$ 
with 
$c_{\rm nst}=c''\int_{0}^{1} dy y^{-1-\theta+\lambda} >0$.
These susceptibilities satisfy the {\it sum rule}
$
%\be
\chi_{\rm ZFC}(t,\tw)+\chi_{\rm
TRM}(t,\tw)=\chi_{\rm ZFC}(t,0)
%\label{eq-sum-rule}
%\ee
$
which must hold for linear responses.

In the special limit of $\tw=0$ we obtain,
\begin{eqnarray}
 \chi_{\rm ZFC}(t,0)
 \sim  \chi_{\rm D} - c'''
\frac{\rhoo m^{2}}{\Upsilon(L(t)/L_{0})^{\theta}}
\label{eq-scaling-chizfc0}
\end{eqnarray}
with $c''' =A(1)-c_{\rm nst}$.
Note that $\chi_{\rm ZFC}(t,0)$  converges to the dynamical
susceptibility $\chid (> \chiea)$ in the limit $t \to \infty$.
This implies  $\chid$ is nothing but the field cooled (FC)
susceptibility $\chi_{\rm FC}$ \cite{cooling rate}, while $\chiea$ is close 
to what is called the ZFC susceptibility $\chi_{\rm ZFC}$. Thus
we expect the well known experimental observation \cite{Uppsala, Saclay}
$\chi_{\rm FC} > \chi_{\rm ZFC}$ 
is a {\it dynamical but not a transient phenomenon} .

The overall feature of the two time quantities so far discussed 
is displayed in Fig. \ref{fig:c-conjecture}.
Note that a parametric plot of $C$ vs $\chi_{\rm ZFC}$
becomes radically different from the conventional picture 
\cite{BCKM,FMGPP} in which the breaking of FDT and TTI 
are supposed to happen simultaneously at $(\qea, \kb T\chi_{\rm EA})$.

%\section{Numerical Results}

\begin{figure}[t]
 \resizebox{\figwidth}{!}{\includegraphics{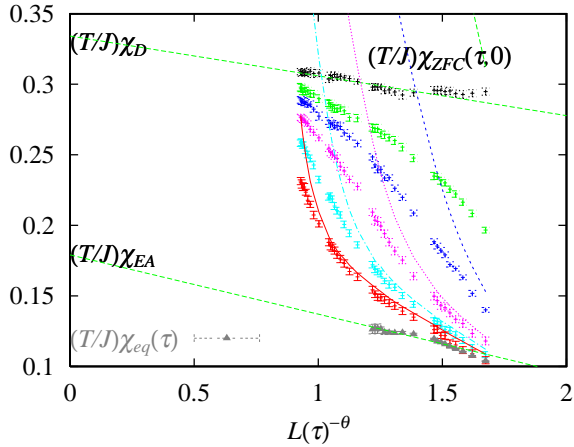}}
 \caption{ZFC  susceptibilities and spin autocorrelation functions 
vs $1/L^{\theta}(\tau)$. 
The symbols are $(T/J)\chi_{\rm ZFC}(\tau+\tw,\tw)$ with
$\tw=0,10,10^{2},10^{3},10^{4},10^{5}$ MC step from the top to the bottom.
The maximum time separation is $\tau=10^{5}$ MC step.
The curves with lines are corresponding $1-C(\tau+\tw,\tw)$.
The data of $(T/J)\chi_{\rm eq}(\tau)$ (filled triangle) 
is shown at the bottom. 
}
\label{fig:tti-fdt-sseparation}
\end{figure}

Finally let us briefly discuss some numerical results concerning 
our scenario. We performed Monte Carlo (MC) simulations of
isothermal aging of a 4 dimensional EA spin-glass model (size $24^{4}$)with 
$\pm J$ interactions ($T_{c}=2.0J$, with $\kb=1$ here
and hereafter)  starting from random initial conditions. 
A set of data of $1-C(\tau+\tw,\tw)$ and $(T/J)\chi_{\rm ZFC}(\tau+\tw,\tw)$
 at $T/J=0.8$ is plotted
against $L(\tau)^{-\theta}$ in Fig. \ref{fig:tti-fdt-sseparation}. 
The field of strength $h/J=0.1$ was
used and the linearity of the responses was checked by the sum rule, 
$\chi_{\rm ZFC}(t,\tw)+\chi_{\rm TRM}(t,\tw)=\chi_{\rm ZFC}(t,0)$.
We used  $\theta=0.82$
obtained by a defect free-energy analysis \cite{KH}.
For the growth law $L(t)$, we used the result of an independent measurement 
of the growth of domain \cite{HYT,YHT}. As explained 
in \cite{KYT,HYT,YHT} the analysis of the quasi-equilibrium 
yields the equilibrium limit curve shown at the bottom of 
the figure (See \cite{HYT}). 
It follows the scaling form \eq{eq-chi-eq} pointing toward 
$(T/J) \chiea \simeq 0.18$. The other extreme $\chi_{\rm ZFC}(\tau,0)$ also
becomes linear in this plot as expected in \eq{eq-scaling-chizfc0}
pointing toward $(T/J) \chid \simeq 0.34$ well above $\chiea$.
Apparently FDT ($(T/J)\chi=1-C$) is well satisfied at small
$\tau$ and broken at large $\tau$. Interestingly enough,
the break points of the FDT move further away from $(T/J)\chi_{\rm eq}(\tau)$
suggesting the separation of the breaking of FDT and TTI as we anticipated.
Such a feature has not been reported
in previous numerical studies \cite{FDT-MC}. 

We also confirmed most of other scaling ansatz 
of the two time quantities within the 4D EA model which 
will be presented elsewhere \cite{YHT}. 
It will be certainly interesting to test our scenario 
by experiments such as noise measurements \cite{HO}
and further numerical simulations.

%\section{Perspective}

\acknowledgments
This work is supported by a Grant-in-Aid for Scientific Research
Program(\#12640367), and that for the Encouragement of Young
Scientists(\#13740233)
from the Ministry of Education, Culture, Sports, Science 
and Technology of Japan.  
The present simulations have been performed on Fujitsu VPP-500/40 at the
Supercomputer Center, Institute for Solid State Physics, the University
of Tokyo.


\begin{thebibliography}{99}

\bibitem{APY-review}%1
 ``Spin Glasses and Random Fields'', 
A.P. Young Editor, (World Scientific,  1998).

\bibitem{Uppsala} P. Nordblad and P. Svedlindh, in \cite{APY-review}.

\bibitem{Saclay}
  E.~Vincent, J.~Hamman, M.~Ocio, J.~-P.~Bouchaud and L.~F.~Cuglinadolo,
in {\it Proceeding of the Sitges Conference on Glass Systems},
 E.~Rubi Editor, (Springer, 1996). 

 \bibitem{Review-Rieger}
H.~Rieger, in {\it Annual Review of Computational Physics II},
	 edited by D.~Stauffer (World Scientific, 1995). 

\bibitem{BCKM}
J. -P. Bouchaud, L. F. Cugliandolo, J. Kurchan
and M. M\'ezard in \cite{APY-review}.


%%%%%%%%% DYNAMICAL MFT             

 \bibitem{MFT}%6
L.~F.~Cugliandolo and J.~Kurchan, 
Phys.~Rev.~Lett. {\bf 71}, 173 (1993); J.~Phys.~A {\bf 27}, 5749 (1994);
S. Franz and M. M\'{e}zard, Europhys. Lett. {\bf 26}, 209 (1994);
Physica {\bf A 209}, 1 (1994). L. F. Cugliandolo and P. Le Doussal,
Phys. Rev. {\bf E 53}, 1525 (1996). 
%L.~F.~Cugliandolo, J.~Kurchan and P. Le~Doussal, Phys. Rev. Lett.
%{\bf 76}, 2390 (1996).

\bibitem{CKP97} L. F. Cugliandolo, J. Kurchan and L. Peliti,
Phys. Rev. E {\bf 55}, 3898 (1997).

%%%%%% DROPLET SCALING THEORY,  DOMAIN GROWTH DYNAMICS

\bibitem{BM} A. J. Bray and M. A. Moore, Phys. Rev. Lett. {\bf 58}, 57 (1987).

 \bibitem{FH1}
D.~S.~Fisher and D.~A.~Huse, Phys.~Rev.~B{\bf 38},  386 (1988). 

 \bibitem{FH2}
D.~S.~Fisher and D.~A.~Huse, Phys.~Rev.~B{\bf 38}, 373 (1988). 

%%%%%%%% SYSTEM SIZE EXCITATIONS

\bibitem{LEE} %11
J. Houdayer and O. C. Martin, Euro. Phys. Lett.
{\bf 49}, 794 (2000);  F. Krazakala and O. C. Martin, Phys. Rev. Lett.
{\bf 85}, 3013 (2000);  M. Palassini and A. P. Young, 
Phys. Rev. Lett. {\bf 85}, 3017 (2000).

\bibitem{NS} C. M. Newman and D. L. Stein, Phys. Rev. {\bf E 57}, 1356
	(1998); Phys. Rev. Lett. {\bf 87}, 077201 (2001).


\bibitem{M99} A. Middleton, Phys. Rev. Lett. {\bf 83}, 1672 (1999).

 \bibitem{KYT}
T.~Komori, H.~Yoshino and H.~Takayama, J. Phys. Soc. Jpn. {\bf 68}, 3387
 (1999);  J.~Phys.~Soc.~Jpn.  {\bf 69}, 1192 (2000). 

\bibitem{low-temp-MC} H. G. Katzgraber, M. Palassini, and A. P. Young, 
Phys. Rev. {\bf B 63}, 184422 (2001) and references there in.

%%%%%%%%%%%%%%%%%%%%%%  NOTE


\bibitem{cooling rate}%16
 Infinitely fast cooling is considered so that TRM
becomes the same as isothermal remanent magnetization (IRM).

\bibitem{note}
Unfortunately the ambiguity of the width $\ln a$ is unavoidable 
within the independent droplet model.

%%%%%%%%%%%%%%%%%%%%%% DOMAIN GROWTH, AGING

\bibitem{B94} A. J. Bray, Adv. Phys. {\bf 43}, 357 (1994).

\bibitem{Uppsala-AC} P. E. J{\"o}nsson, H. Yoshino, P. Nordblad,
H. Aruga Katori and A. Ito, cond-mat/0112389.

\bibitem{Saclay-AC} V. Dupis, E. Vincent, J.-P Bouchaud, J. Hammann, A,
Ito and H. A. Katori., Phys. Rev. {\bf B} 64 174204 (2001).

\bibitem{HYT}
K.~Hukushima, H.~Yoshino and H.~Takayama, 
Prog. Theor. Phys. Supp. {\bf 138}, 568 (2000), cond-mat/9910414. 

\bibitem{YHT}
H.~Yoshino, K.~Hukushima  and H.~Takayama, cond-mat/0202110.

\bibitem{BB}
L. Bertheir and J.-P. Bouchaud, cond-mat/0202069.

\bibitem{FMGPP} %21
S. Franz, M. M\'{e}zard, G. Parisi, L. Peliti,
 Phys. Rev. Lett. {\bf 81}, 1758  (1998).

%%%%% DEFECT FREE=ENERGY

\bibitem{KH}
K.~Hukushima, Phys.~Rev.~E {\bf 60}, 3606 (1999). 

%%%%% PREVIOUS MC

\bibitem{FDT-MC} S. Franz and H. Rieger, J. Stat. Phys. {\bf 79}, 749 (1995);
G.~Parisi, F.~Ricci-Tersenghi and J.~J.~Ruiz-Lorenzo, 
J.~Phys.~ A {\bf 29}, 7943 (1996); 
E.~Marinari, G.~Parisi, F.~Ricci-Tersenghi and J.~J.~Ruiz-Lorenzo, 
J.~Phys.~A {\bf 31}, 2611 (1998).

%%%%% NOISE

\bibitem{HO} D. H\'{e}risson and M. Ocio, 
cond-mat/0112378.

\end{thebibliography}
\end{document}